\documentclass[12pt]{article}
\usepackage{graphicx}
\usepackage{amsmath}

\begin{document}


\vspace*{2cm}

\begin{center}
{\bf\Large Geometry of historical epoch,
\\
the Alexandrov's problem

\medskip

and  non-G\"odel quantum time machine}

\vspace*{1cm}

Alexander K. Guts
\vspace*{0.6cm}

 Dostoevsky Omsk State University,
\medskip

Department of Computer Sciences,
\medskip

pr.~Mira, 55a, Omsk, 644077, Russia
\medskip

 Email: aguts@mail.ru

 November 6, 2016.

\end{center}

\vspace*{0.6cm}\noindent \textbf{Abstract}. The new quantum
principle of a time machine that is not using a smooth timelike
loops in Lorentz manifolds is described. The proposed time machine
is based on the destruction of interference of quantum
superposition states in the Wheeler superspace.


\begin{center}

\vspace*{3cm} {\small The version of the article "Non-G\"odel time
machine"\\ that is published in the Russian journal\\ Mathematical
Structures and Modelling. 2016. No. 3 (39). P.48-58.}
\end{center}


\def\C{{{\rm I}\!\hspace*{-1.3mm}{\rm C}}}
\def\R{{{\rm I} \! {\rm R}}}
\def\g{{\cal G}}
\def\E{{\cal E}}
\def\t{{ {\scriptstyle {}^{(3)}\hspace*{-0.45mm} }}}
\def\f{{ {\scriptstyle {}^{(4)}\hspace*{-0.45mm} }}}

\newpage

 It is known \cite {guc1,guc1a}, that the classical space-time
$M^4$
 is the result of quantum interference occurring in the Wheeler's
 superspace due to  quantum superposition
 \begin{equation}\label{packet1}
\Psi(M^4) = \sum_{k\in K} c_k\Psi(\Omega_k), \ \ \ c_k \in {{\rm
I} \! \hspace*{-1.3mm} {\rm C}},
\end{equation}
$$
\Psi(\Omega_k) = A_k \left(\begin{array}{l}
  \mbox{\footnotesize slowly varying} \\
\mbox{\footnotesize amplitude function}
\end{array}\right)e^{-\frac{i}{\hbar}S_k(\t\g)},
$$
$$
S_k = S_{k^\prime} = ... = S,
$$
where a wave function $\Psi(\Omega_k) = \Psi_k(\t\g) $ is a
functional of 3-dimensional Riemannian geometry $\t\g = (M^3,
h_{\alpha\beta}) $ and satisfies the functional Wheeler-DeWitt's
equation:
$$
 \left(G_{\alpha\beta\gamma\delta}\frac{\delta }{\delta
h_{\alpha\beta}}\frac{\delta }{\delta h_{\gamma\delta}}+\sqrt{h}\t
R
    -\E(h_{\alpha\beta},\mu,B,e,\sigma,\nu)\right)\Psi_k(\t\g)=0.
$$
 Here $S_k(\t\g)$ is an action
that satisfies to the Einstein-Hamilton-Jacobi equation
$$
G_{\alpha\beta\gamma\delta}\left(\frac{\delta S_k}{\delta
h_{\alpha\beta}}\right)\left(\frac{\delta S_k}{\delta
    h_{\gamma\delta}}\right)-\sqrt{h}\ \t R+\E(h_{\alpha\beta},\mu,B,e,\sigma,\nu)=0.
$$

\vspace*{6mm} \noindent {\bf \large 1. A historical epoch}
\medskip

Usually, one does not discuss the meaning of the system $\Omega$
and its states $\Omega_k, \ k \in K,$ which have the wave function
$\Psi(\Omega_k)$. In \cite{guc2} we suggested that  state
$\Omega_k$ is a {\it real} historical epoch, which is relatively
stable, unchanging, constant in all senses including the geometry
of 3-space, i.e. the stable time period of the existence of human
civilization. These historical epoches "without time" were
described as Gestalt by Wolfgang Goethe and specially Oswald
Spengler in "Der Untergang des Abendlandes".

The 3-geometry $\t\g$ of historical epoch $\Omega_k$ {\it knows}
its {\it time location} in 4-geometry of space-time: "the
hypersurface drawn through spacetime to give one $\t\g$ can be
pushed forward in time a little here or a little there or a little
somewhere else to give one or another or another new $\t\g$. {\it
"Time}" conceived in these terms {\it means} nothing more or less
than {\it the location of the $\t\g$ in the} $\f\g$. In this sense
3-geometry is a carrier of information about time"
\cite[p.\,37]{guc1a}.

Thus, we can say that space-time is the result of the quantum
interference of historical epochs. The Minkowski postulate of {\it
absolute World of events}, or absolute space-time should be
replaced by following postulate: {\it The World exists in the form
of historical epoches}. So, we can make the transition in each
historical epoch.

The most important fact of quantum theory is observation, and that
the interference pattern disappears when one trying to measure the
state of quantum system $\Omega$. In our case, this means that the
beginning of  procedure of geometrical measurement, i.e.
 switching on of apparatus that fix the particular geometry
$\t\g^\prime$, which took place in the past, one inevitably
destroys {\ it locally} interference pattern.

In fact, let $|k\rangle$ be a state of historical epoch
$\Omega_{k}$ with wave function $\Psi_k(\t\g)$ of quantum system
$\Omega$, which is described by superposition
$$
\sum_{k}c_k|k\rangle.
$$
Let an observer $X$ is living in epoch $\Omega_{\alpha_0}$. Then
geometrical measurement that is produced by observer $X$ with the
help of  apparatus ${\cal A}$ with initial state $A$ and  aimed at
specific value of 3-geometry $\t\g^\prime$ which is geometry of
past historical epoch $\Omega_{k^\prime}$, give the new
superposition:
$$
\left(\sum_k c_k|k\rangle\right)\otimes|A\rangle\to \left(\sum_{k,
k\neq k_0,
k^\prime}c_k|k\rangle\right)\otimes|\widetilde{A}\rangle +
\left(c_{k_0}|k_0\rangle\otimes
|A_0\rangle+c_{k^\prime}|k^\prime\rangle\otimes
|A^\prime\rangle\right).
$$
We see that two epoches $\Omega_{k_0}, \Omega_{k^\prime}$ form the
entangled states with apparatus (environment). The others epoches
form  the interference
 quantum superposition, giving classical Universe.

Historical epoches exist {\it  at the same time}, or {\it
simultaneously}. External observers are not out of this quantum
superposition, and within own historical epoch for which this own
historical epoch seems true objective reality. From the epoch
$\Omega_{k_0}$ the observer $X$ makes the observation of packet
(\ref{packet1}). This packet makes collapse and some spatial
volume of $\Omega_{k^\prime}$ is localized in the epoch $
\Omega_{k_0}$. As a result, apparatus together with the observer
$X$ will be in Reality which is past historical epoch
$\Omega_{k^\prime}$, i.e. apparatus together with the observer
will enter to past. In other words, we have mechanism called  the
{\it time machine}.

\vspace*{6mm} \noindent {\bf \large 2. The G\"odel time machine}
\medskip

In the General relativity the time machine refers to the return
mechanism in the past, suggested in 1949 famous logician Kurt
G\"odel. The {\it G\"odel time machine } suggests that the
transition in the past makes the apparatus, world line of which is
a closed time-like smooth curve.

In 1968 Soviet academician A.D.~Alexandrov formulated  the the
problem of founding of the physical conditions under which
 one is possible to transition in a past historical epoch \cite{guc3,
guc4}. But he considered  then the only known mechanism of the
G\"odel time machine. Therefore,  the  Alexandrov problem is
limited the study of Lorentz manifolds in which exist a closed
smooth time-like curves.

Situation with solving of the Alexandrov problem is given in
\cite{guc4,guc4a}.

It was natural to try to leave the General relativity in its
classic version, i.e. within the framework of classical physics,
and to use the ideas of quantum theory. The idea of using quantum
interference for transitions in time was first described in the
article \cite{guc5}.

\vspace*{6mm} \noindent {\bf \large 3. The non-G\"odel quantum
time machine}
\medskip

Our project that  we set out at the beginning of this note is also
a quantum time machine. Described the transition in the past is
probabilistic. The required historical epoch  can not be achieved
at all times. The full description of this project to be soon
published in \cite{guc6}.

The transition in past epoch is done by measuring of the geometry
(and not only geometry, because quantum theory also uses
observers, i.e. the moments of human consciousness).  We
understand very well how to measure the geometry. Within the
ideology of the General relativity the measurement of geometry is
the measurement of gravitational fields. Every historical epoch
can be regarded as a gravitational wave with fixed parameters,
detection apparatus of which is more or less developed.

A certain kind of activity (measurement) of observers from epoch
$\Omega_{k_0}$ with respect to values, that are typical only for
the epoch $\Omega_ {k^\prime},$ destroy the entanglement with the
last remaining epochs, and is linked these two epoches. This
measurement localizes  each epoch into another. There will be a
transition from epoch to another. This is nothing other than a
time machine. Since another epoch have my quantum double (this is
as another location of one particle), then murder him  does not
mean murder themselves. In other words, the grandfather paradox is
solved trivially.

But the problem of measurement of geometry is not easy one.
Wheleer wrote: "The formalism of quantum gravity, in its best
developed form, makes 3-geometry a central concept; consequently,
the finding of a proper way to measure this 3-geometry -- against
the background of the indeterminancy of the conjugate
geometrodynamic field momentum -- is a central issue"
\cite[p.\,231]{wheeler1}.

 Since $|\Psi(\Omega_k)|^2$ is the
probability density, then we must assume that historical epoch has
the most probable geometry of ${\t\g}$. For the fixing of specific
geometry, one should be adjusted instrumentation. Perhaps it is
not just measuring apparatus, but devices  which give the
"resonance" with geometry of the historical epoch that we are
interested through the generation of gravitational waves {\it or
others fields} \cite{Zeh1986,Joos1986}.

But "tuning equipment" to the desired geometry, perhaps is only
first step. Quantum mechanics is strongly related to the
consciousness problem \cite{guc7}, and  historical epochs are not
empty geometric worlds, but are worlds inhabited by people, and
therefore historical epoch is also the energy of ethnic groups,
political passion, and more. The geometry is need  people to live,
and  quantum interference patterns
 are destroyed not only
due to the measurements, but also because of the intentions to
make  such measurements.

\vspace*{6mm} \noindent {\bf \large 4.  The quantum time machine
in minisuperspace}
\medskip

The Wheeler-DeWitt equation for minisuperspace which is defined by
a homogeneous and isotropic Universe (compact spatial topology, $k
= 1$)
$$
ds^2=N^2(t)dt^2-a^2(t)dl^2_{S^3}
$$
with a massless scalar field $\phi$ and $a=e^\alpha, \
\alpha\in\R,$ has the following form
\begin{equation}\label{wdw}
\left(\frac{\partial^2}{\partial^2
\alpha}-\frac{\partial^2}{\partial^2
\phi}-e^{4\alpha}\right)\Psi(\alpha,\phi)=0.
\end{equation}
 We impose the
boundary condition
\begin{equation}\label{wdw1}
\lim_{\alpha\to 0}\Psi(\alpha,\phi) = 0
\end{equation}
 necessary in order to reconstruct
semiclassical states describing the dynamics of a closed Universe,
since regions of minisuperspace corresponding to arbitrary large
scale factors are not accessible.

It should be noted that the considered scalar field $\phi$  most
likely to be regarded not as matter, but as a collective energy of
people which create the universe \cite{guc8,guc9}.

The problem (\ref{wdw})--(\ref{wdw1}) has the following solution
\cite{guc10} in the form of wave packet
\begin{equation}\label{wp}
\Psi(\alpha,\phi) = \int\limits_{-\infty}^{+\infty}A(k)
\Psi_k(\alpha,\phi)dk,
\end{equation}
where
\begin{equation}\label{wdw8}
\Psi_k(\alpha,\phi) =e^{-k\frac{\pi}{4}}\sqrt{k} K_{ik/2} \left(
\frac{e^{2\alpha}}{2}\right)e^{ik\phi},
\end{equation}
\begin{equation}\label{wdw9}
A(k)= \frac{1}{\pi^{1/4}\sqrt{b}} e^{-\frac{(k-\bar{k})^2}{2b^2}}.
\end{equation}

\begin{figure}[h]
\centering\includegraphics[height=5.5cm]{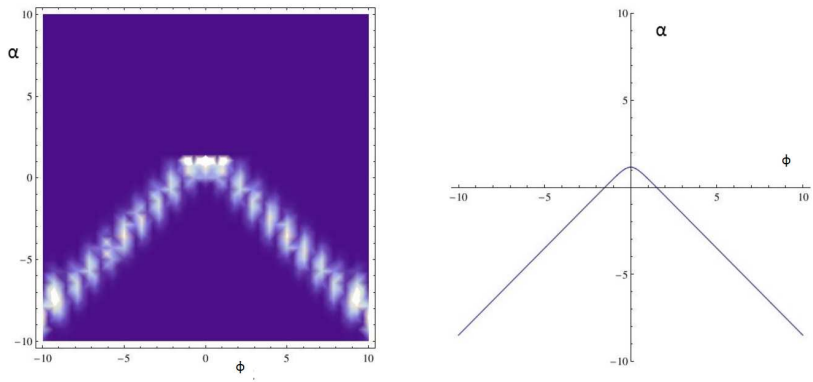}
\vspace{-0.2cm}

\medskip
\caption{\footnotesize a) Absolute square of the “wave function”
of the Universe corresponding to the choice of parameters $b = 1,
\bar{k} = 10$. Lighter shades correspond to larger values of the
wave function; b) Classical trajectory of the Universe in
minisuperspace. It is closely followed by the peak of
semiclassical states, as we can see by comparison with a).
\cite{guc10}}\label{pic1}

\end{figure}

The solution represents a wave packet that starts propagating from
a region where the potential vanishes (i.e. at the initial
singularity) towards the potential barrier located approximately
at $\alpha_{\bar{k}} = \frac{1}{2} \log \bar{k}$, from where it is
reflected back. For such values of $\bar{k}$ the wave packet is
practically completely reflected back from the barrier. The
parameter $b$ gives a measure of the semiclassicality of
 the state,
i.e. it accounts for how much it peaks on the classical
trajectory. We remark that the peak of the wave packet follows
closely the classical trajectory,
\begin{equation}\label{wdw10}
e^{2\alpha} =  \frac{\bar{k}}{ch(2\phi)}.
\end{equation}
 and refer the reader to Fig.~\ref{pic1}, b).

\begin{figure}[h]
\centering\includegraphics[height=9cm]{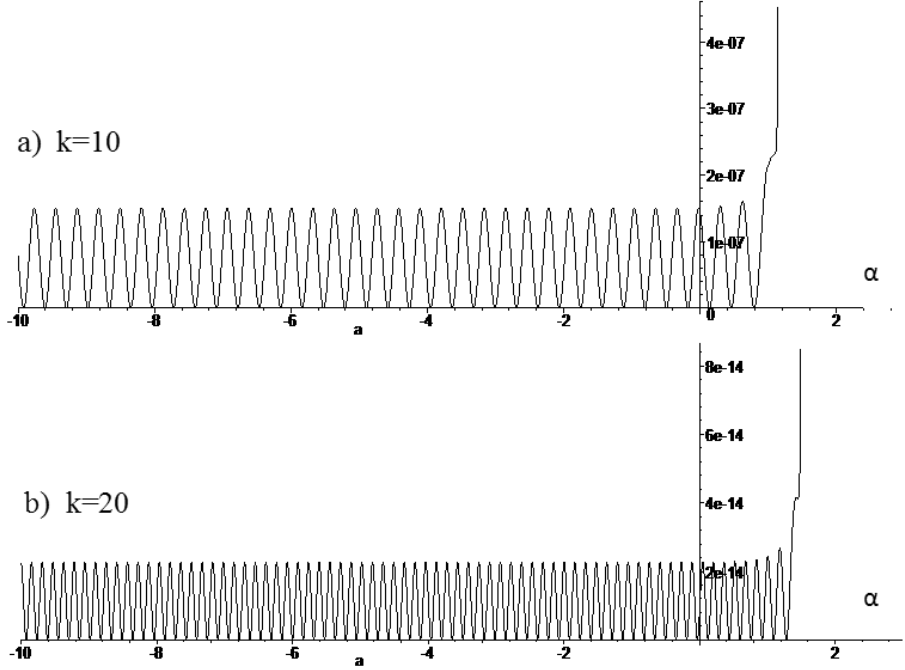}
\vspace{-0.2cm}

\medskip
\caption{\footnotesize Graph of $|\Psi_{k}(\alpha,\phi)|^2$ for
$k=10$  and $k=10$.}\label{epoch}

\end{figure}

Under mesuarement we have a collapse of considered wave packet
$$
\int\limits_{-\infty}^{+\infty}A(k) \Psi_k(\alpha,\phi)dk \to
\Psi_{k}(\alpha,\phi)
$$
with probability $|A(k_0)|^2$.

 Consider the graphs at Fig.~\ref{epoch} of geometric epoches
$\Omega_{k_0}$ for $k_0=10$ and $k_0=10$.

We see that historical epoch $\Omega_{k_0}$ can have a number of
different geometries.  But a  geometry with $\alpha
>0.4$  will be most probable for $\Omega_{10}$ and with $\alpha
>0.6$ for $\Omega_{20}$. But what is important: geometry $\alpha <0.4$
is much less likely for $\Omega_{20}$ than for $\Omega_{10}$.
Although transition to geometry with $\alpha <0.4$ is practically
impossible. The differences in the behavior of the probability
density functions of various historical epoches give the
opportunity to adjust the measuring equipment so that to get to
the desired epoch.

\end{document}